\documentclass[runningheads]{llncs}
\usepackage[T1]{fontenc}
\usepackage{graphicx}
\usepackage{mathpartir}
\usepackage{amsmath}
\usepackage{orcidlink}
\usepackage{subcaption}
\usepackage{listings}
\usepackage{newfloat}        
\DeclareFloatingEnvironment[fileext=lol]{listing}

\lstdefinelanguage{Go}{
  morekeywords={break,case,chan,const,continue,default,defer,else,
    fallthrough,for,func,go,goto,if,import,interface,map,package,
    range,return,select,struct,switch,type,var},
  morekeywords=[2]{bool,byte,error,int,string,true,false,nil,make,close},
  sensitive=true,
  morecomment=[l]{//},
  morecomment=[s]{/*}{*/},
  morestring=[b]"
}
\lstdefinestyle{gostyle}{
  language=Go,
  basicstyle=\ttfamily\footnotesize,
  keywordstyle=\color[HTML]{377E21}\bfseries,
  numbers=left, numberstyle=\tiny, numbersep=3pt,
  breaklines=true,
  tabsize=2,
  escapeinside={|}{|},
  columns=flexible, keepspaces=true, showstringspaces=false,
  aboveskip=0pt, belowskip=0pt
}
\lstnewenvironment{minted}[2][]{\lstset{style=gostyle}}{}

\DeclareCaptionSubType{listing} 

\usepackage{xurl}
\usepackage{hyperref}
\usepackage[htt]{hyphenat}
\usepackage{cleveref}
\usepackage{booktabs}
\usepackage{makecell}
\usepackage[disable]{todonotes}
\usepackage{proof}
\usepackage[table,svgnames]{xcolor}
\usepackage{color}

\usepackage{xspace}
\usepackage{pifont}

\usepackage[inline]{enumitem}

\AddToHook{env/minted/after}{\vspace{-13pt}}

\newcommand{\cmark}{\textcolor{green!70!black}{\ding{51}}\xspace}
\newcommand{\xmark}{\textcolor{red}{\ding{55}}\xspace}
\newcommand{\GDDR}{\textsc{GoDDaR}\xspace}
\newcommand{\NeoGDDR}{\GDDR}
\newcommand{\migoinf}{\textit{migoinfer}\xspace}
\newcommand{\goat}{\textsc{goat}\xspace}
\newcommand{\migo}{\textsc{MiGo}\xspace}
\newcommand{\arr}[1]{\mathrel{\stackrel{#1}{\longrightarrow}}}
\newcommand{\rname}[1]{[\textsc{#1}]\xspace}
\newcommand{\co}[1]{\ensuremath{\overline{#1}}}

\newcommand{\pars}{\ensuremath{\mathbin{\mid}}}
\newcommand{\inert}{\ensuremath{\mathbf{0}}\xspace}

\newcommand{\prf}[1]{\ensuremath{#1.}}
\newcommand{\coprf}[1]{\prf{\bar{#1}}}

\newcommand{\paral}{\ensuremath{\!\parallel\!}}
\newcommand{\paralS}{\ensuremath{\parallel}}

\newcommand{\deftxt}{\stackrel{\textbf{\tiny def}}{=}}

\newcommand{\env}{\ensuremath{\Gamma}\xspace}

\newcommand{\mi}{\ensuremath{\downarrow}}
\newcommand{\mo}{\ensuremath{\uparrow}}

\newcommand{\dentOalt}[2]{\textbf{dr1}\text{(}\ensuremath{#1,\ #2}\text{)}}
\newcommand{\dentTalt}[2]{\textbf{dr2}\text{(}\ensuremath{#1,\ #2}\text{)}}
\newcommand{\added}[1]{\textcolor{cyan!70!black}{#1}}
\newcommand{\removed}[1]{\textcolor{red}{#1}}
\begin{document}
\title{Automatically Detecting and Fixing Deadlocks in Go Code with \NeoGDDR
}
\titlerunning{Automatically Detecting and Fixing Deadlocks in Go Code with \NeoGDDR}
%
\author{Gustavo Feio\orcidlink{0009-0008-0802-1005} \and
   Jorge Coelho \and
   Miguel Flor\orcidlink{0009-0003-9535-5970} \and
   João M. Lourenço\orcidlink{0000-0002-8495-6442} \and
   António Ravara\orcidlink{0000-0001-8074-0380}
}
\authorrunning{G. Feio et al.}
%
\institute{NOVA LINCS and NOVA FCT, Portugal}
\maketitle
%
\begin{abstract}
The Go programming language provides a lightweight and accessible abstraction for concurrent programming through goroutines, which are prone to deadlocks.
These are notable concurrency bugs, which can prevent the progress of two or more threads, a situation that is difficult to detect and reproduce.
To mitigate deadlocks, the Go programming language includes a runtime detector which aborts execution if all threads are blocked and the whole program cannot progress (reached a \emph{global deadlock}).
However, due to the inherently nondeterministic nature of thread scheduling, this runtime mechanism can only detect global deadlocks that have already manifested during execution.
In addition, the runtime detector is unable to identify \emph{partial deadlocks}, where a subset of goroutines is permanently blocked while at least one remains runnable.
The ability to statically detect deadlocks, namely local ones, is essential for the development of dependable concurrent software.

While several tools have been proposed to statically detect deadlocks in concurrent programs, only a limited number address the equally important problem of assisting developers in fixing them.
Detecting and resolving partial deadlocks requires reasoning about complex interleavings and communication patterns, an inherently challenging task.

In this paper, we present \NeoGDDR, a tool that not only detects unobserved global or partial deadlocks in Go programs, but crucially, also suggests concrete fixes.
\NeoGDDR translates Go source code into a rich intermediate representation that reflects the message-passing communication events and their composition patterns.
Partial deadlocks are identified through symbolic analysis over this representation.
To resolve the detected deadlocks, we implemented algorithms that transform the intermediate representation to eliminate the problematic synchronisation patterns.
From this enriched representation, \NeoGDDR generates fix suggestions for the original Go source code in the form of a \emph{diff}.

We evaluate \NeoGDDR on a benchmark suite collecting representative deadlock examples from the literature and git repositories.
Our evaluation has two goals: validating the correctness of the proposed algorithms and comparing their performance against the state of the art.
The results demonstrate that \NeoGDDR is competitive with the best existing approaches in partial deadlock detection and advances the state of the art in automated deadlock fixing.

\end{abstract}

\section{Introduction}
Go is a statically typed programming language featuring native support for concurrency via lightweight execution units: the goroutines. Improper synchronisation of these units introduces vulnerabilities to concurrency-related defects. To mitigate these risks, the language design prioritises channel-based communication, adhering to the principle of sharing memory by communicating rather than communicating by sharing memory~\cite{GoProverb,EffectiveGo}. This channel-based communication mechanism facilitates the prevention of data races by enforcing a formal transfer of data ownership between concurrent processes.
Despite these efforts, threads may begin to block if they try acquiring a resource belonging to another goroutine that will not relinquish its ownership, and once enough threads fall into a blocking state, the program as a whole will block.
This effect where threads start blocking due to unsatisfiable dependencies is known as \emph{deadlocking}, and it can be categorised as \emph{local} (or \emph{partial}), where only some threads (not all) are blocked, or \emph{global} (or \emph{total}), where all threads have blocked and, thus, the whole program as well.
If a program falls into a global deadlock, Go's runtime deadlock detector will kick in and abort it, indicating where the program deadlocked and the stack trace leading up to that point.

While the runtime deadlock detection is helpful, it only applies to observed global deadlocks, and it would be better to detect the blocking scenarios before even executing the program, in order to save both time and resources.
Due to this necessity, many static deadlock detecting tools have been researched and developed, showing promising results with reasonable accuracy on bug-ridden programs.
Static deadlock detectors have greatly improved the accessibility of concurrent program development, but there is still the need to solve any detected bug---a task easier said than done, since real-world programs tend to become very complex, especially in a concurrent context.
Complementary to deadlock detection, deadlock resolution is still under-researched, with few tools addressing the challenge. We build on the approach of Francalanza \emph{et al.}~\cite{FrancalanzaGR15}, which provides an algorithm to disentangle deadlocks in a linear version of CCS~\cite{MilnerCCS}, and show how it can be used to effectively propose fixes to deadlocked Go code.

To bridge the gap between static detection and deadlock resolution, this paper proposes \NeoGDDR. By utilising intermediate representations of the source program, \NeoGDDR employs symbolic analysis to identify deadlock scenarios. The tool subsequently derives resolution strategies and outputs these modifications to the user as standardised \emph{diffs} applicable to the original Go source code.
For instance, consider two channels $a$ and $b$, and two threads $t_1$ and $t_2$ which communicate through those channels.
It is remarkably easy for a programmer to mistakenly order some input and output actions on those channels such that the two threads become mutually dependent, resulting in a deadlock: let $t_1$ read from $a$ and then write on $b$, and $t_2$ read from $b$ and then write on $a$; since actions on channels must synchronise, and $t_1$ is waiting on $a$, and $t_2$ is waiting on~$b$, the program enters a deadlocking state.
\NeoGDDR detects the deadlock and generates a \emph{diff} containing a synchronisation fix. This modification parallelises the blocking operation, resolving the deadlock and restoring program execution.

To evaluate \NeoGDDR performance, the tool was benchmarked against state-of-the-art detection frameworks using a curated collection of deadlock instances~\cite{BugCollection}. Notice that we focus on deadlocks caused by incorrect channel synchronisation---we do not consider deadlocks resulting from incorrect use of locks to ensure access to shared memory in mutual exclusion. The tool only analyses the channel-based communication structure of the Go code.
Furthermore, the generated resolution strategies were evaluated through a comparative analysis with fixes produced by existing automated tools.
Experimental results indicate that \NeoGDDR exceeds the detection accuracy of existing alternatives in deadlock detection. The tool generates viable resolution strategies that surpass the performance of the empirical methods used by the current state-of-the-art tools, which exhibit more limited applicability.

\section{Related Work}
Dynamic deadlock detection in Go make use of methodologies for runtime monitoring. Golf~\cite{Golf} integrates with the Go garbage collector to approximate liveness during the marking phase, identifying deadlocked goroutines when they depend on resources unreachable by other execution units. GoLeak~\cite{GoLeak_LeakProf} detects deadlocks by enforcing an invariant where no goroutines remain after unit test completion. LeakProf~\cite{GoLeak_LeakProf,LeakProf} performs periodic stack trace sampling to detect deadlocks in long-running applications that do not manifest during unit testing. These implementations prioritise detection over completeness.


However, often the objective is to prevent deadlocks from ever occurring. Static analysis facilitates deadlock prevention by verifying program properties without execution. Nano-Go~\cite{nanoGo} and Mini-Go~\cite{MiniGo_Automata} model Go language subsets using communication-based representations to perform static property checks. While not specifically aimed at deadlocks, Go2Pins~\cite{Go2Pins} provides a framework for formal verification using Linear Temporal Logic (LTL) formulae, enabling the specification of liveness and safety properties to identify deadlocks. 

These tools illustrate the trade-off between static and dynamic methodologies. Dynamic deadlock detection tools offer precision by observing runtime behaviour, but are limited to specific execution paths. Static approaches enable detection prior to execution, yet rely on abstractions or language subsets that impact model fidelity or increase computational complexity.


\section{\NeoGDDR}
\begin{figure}[t]
    \centering
    \includegraphics[width=0.8\linewidth]{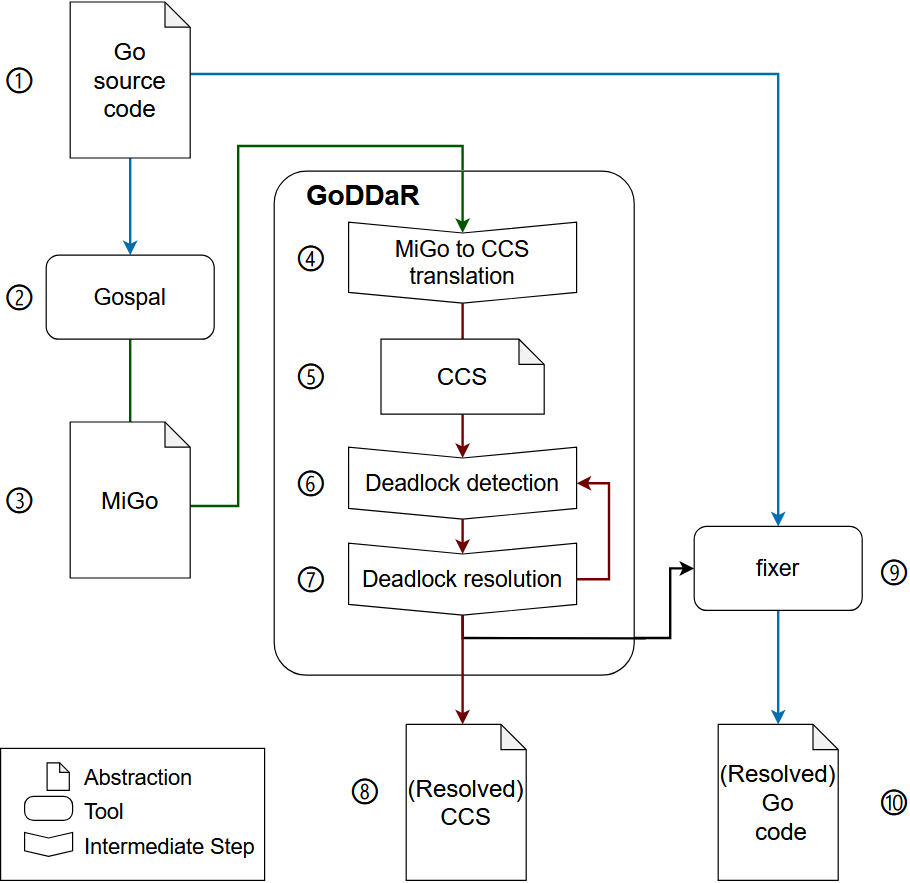}
    \caption{\NeoGDDR pipeline}
    \label{fig:pipeline}
\end{figure}

\NeoGDDR~\cite{GoDDaR}, Go Deadlock Detection and Resolution, is our take on static deadlock detection in Go programs.
The tool is written in OCaml and Go, and is capable of performing symbolic analysis on the Calculus of Communicating Systems (CCS)~\cite{MilnerCCS},
a representation obtained by translating \migo~\cite{MiGo} programs inferred from Go with the \migoinf tool, available through Gospal~\cite{Gospal}.
The general pipeline of \NeoGDDR analysis is illustrated in~\Cref{fig:pipeline}.

\subsection{CCS with Internal and External Choice}
While Milner's CCS~\cite{MilnerCCS} effectively models concurrent processes, Go's semantics include features that standard CCS cannot fully capture. Specifically, representing the Go select statement as an internal (non-deterministic) choice is inaccurate; internal choice implies any branch may be selected regardless of the environment's state, whereas a select branch in Go only executes if its communication can proceed.

To model idiomatic Go accurately, we adopt a version of CCS featuring external choice, following De Nicola and Hennessy~\cite{Nicola1987}. In this calculus, a process reduces to a specific branch only upon synchronisation with an external action. This mechanism over action-prefixed operations correctly encodes the behaviour of the select statement. Additionally, we include an input-prefixed iterative constructor to model unbounded computation.

The language syntax, defined over a countable set of action names $a,b,\ldots$, is in~\Cref{fig:ccs_syntax}, with its operational semantics in~\Cref{fig:CCS_SOS}. Regarding action names as channels, $a$ denotes an input endpoint and $\overline a$ denotes an output endpoint.


\begin{figure}[tb]
\begin{align*}
    \alpha &::= \tau~|~a~|~\overline{a} \\
    P &::= \alpha.P~|~a^\ast.P~|~(P + P)~|~(P~\&~P)~|~(P~|~P)
\end{align*}
    \caption{Syntax of CCS with internal and external choice.}
    \label{fig:ccs_syntax}
\end{figure}

\begin{figure}[tb]
        \centering
        \[
        \begin{array}{l@{\hskip 3em}l}
                        \infer[\rname{Pre}]
                        {\alpha.P \arr{\alpha} P}
                        {}
                &
                        \infer[\rname{L-IntSum}]
                        {(P + Q) \arr{\tau} P'}
                        {}
                \\[1em]
                        \infer[\rname{L-Par}]
                        { (P \pars Q)\arr{\alpha} (P' \pars Q)}
                        { P \arr{\alpha} P'}
                &
                        \infer[\rname{L-Sync}]
                        {(P\pars Q)\arr{\tau}  (P'\pars Q')}
                        {P\arr{a} P' \quad  Q\arr{\co{ a }} Q'}
                \\[1em]
    \infer[\rname{L-ExSum}]
    {(P \mathbin{\&} Q) \arr{\alpha} P'}
    {P \arr{\alpha} P'}
    &
                        \infer[\rname{L-Repl}]
                        { (a^\ast.P \pars  Q) \arr{\tau}  ((a^\ast.P \pars P ) \pars Q') }
                        { Q \arr{ \co{a} } Q'}
        \end{array}
        \]
        \caption{Structural Operational Semantics (SOS) of CCS. Rules IntSum, Par, Sync, ExSum and Repl also have right rules.}
        \label{fig:CCS_SOS}
\end{figure}

\subsection{Deadlock Detection Algorithm}\label{sec:detect_alg}

The implemented detection algorithm is an extension of that of Francalanza et al.~\cite{FrancalanzaGR15} for a finitary linear CCS. We briefly explain it here, but a detailed presentation can be found elsewhere~\cite{GoDDaR}. In a nutshell, the idea is to pair complementary actions within the CCS process to enable state transitions. 
For instance, an output on a channel must synchronise with a corresponding input for both to proceed. 
When matched, these actions undergo a \emph{reduction} (a $\tau$ transition), and the algorithm evaluates the state of the resulting process. 
\NeoGDDR exhaustively explores all execution paths, and a process is considered deadlocked if no further reductions are possible while unsynchronised actions remain. 
Because any blocked action is identified regardless of its thread of origin, this approach detects local deadlocks. 
The following example shows a reduction to a circular deadlock.
\begin{gather}
\prf{\bar{a}}{\prf{b}{\prf{\bar{c}}{\inert}}}~|~\prf{a}{\prf{c}{\prf{\bar{b}}{\inert}}}\label{ccs:fst} \\
    {\prf{b}{\prf{\bar{c}}{\inert}}}~|~{\prf{c}{\prf{\bar{b}}{\inert}}}\label{ccs:snd}
\end{gather}
In the first iteration of the process~\eqref{ccs:fst}, the algorithm matches the actions $a$ and $\bar{a}$ and proceeds with the analysis. 
In process~\eqref{ccs:snd}, however, the available actions $b$ and $c$ cannot synchronise. 
As no further reductions are possible, the process reaches a deadlock. 
\NeoGDDR reports the deadlock and provides the execution trace leading to that state. 
A simpler case is illustrated below.

\subsection{Deadlock Resolution Algorithms}
During analysis, \NeoGDDR reports detected deadlocks and their associated CCS expressions, optionally providing suggested fixes as \textit{diffs}. The resolution phase isolates and parallelises problematic actions using two heuristics:

\begin{itemize}
    \item The first refactors problematic actions into a parallel composition.
    \item The second targets communication primitives: it applies the first heuristic to outputs, while for inputs, it synthesises and parallelises a complementary output action.
\end{itemize}

These heuristics are defined as recursive functions, homomorphic over all operators except prefixing, as shown in~\Cref{fig:disentangling-1,fig:disentangling-2}.

\begin{figure*}[tb]
\begin{minipage}[b]{0.47\linewidth}
    \begin{equation*}
      \begin{aligned}
      \dentOalt{\env\!}{\!\prf{a}{P}}        &\deftxt
                                             \begin{cases}
                                                  \prf{a}{\inert}\paral P &   \env(a) =\,\mi\\
                                                  \prf{a}{(\dentOalt{\env\!}{\!P})}  & \text{otherwise}
                                             \end{cases} \\[1ex]
      \dentOalt{\env\!}{\!\coprf{a}{P}}      &\deftxt
                                             \begin{cases}
                                                  \coprf{a}{\inert}\paral P &   \env(a) =\,\mo\\
                                                  \coprf{a}{(\dentOalt{\env\!}{\!P})}  & \text{otherwise}
                                             \end{cases}
    \end{aligned}
  \end{equation*}
  \caption{First deadlock resolution algorithm. Prefix rules.}
  \label{fig:disentangling-1}
\end{minipage}
\hfill
\begin{minipage}[b]{0.47\linewidth}
  \begin{equation*}
    \begin{aligned}
      \dentTalt{\env\!}{\!\prf{a}{P}}        &\deftxt
                                             \begin{cases}
                                                  \prf{a}{(\dentTalt{\env\!}{\!P})} \paral \coprf{a}{\inert} &   \env(a) =\,\mi\\
                                                  \prf{a}{(\dentTalt{\env\!}{\!P})}  & \text{otherwise}
                                             \end{cases} \\[1ex]
      \dentTalt{\env\!}{\!\coprf{a}{P}}                 &\deftxt
                                             \begin{cases}
                                                  \coprf{a}{\inert}\paral (\dentTalt{\env\!}{\!P}) &   \env(a) =\,\mo\\
                                                  (\dentTalt{\env\!}{\!P}) &   \env(a) =\,\mi\\
                                                  \coprf{a}{(\dentTalt{\env\!}{\!P})}  & \text{otherwise}
                                             \end{cases}
    \end{aligned}
  \end{equation*}
  \caption{Second deadlock resolution algorithm. Prefix rules.}
  \label{fig:disentangling-2}    
\end{minipage}
\end{figure*}
\Cref{lst:resolution_example_go} illustrates a deadlock resulting from incorrect operation ordering; goroutines block at lines~\ref{lst:resolution_example_go_dl1} and~\ref{lst:resolution_example_go_dl2}. The corresponding CCS process is shown in~\Cref{lst:resolution_example_CCS}.
%
\begin{figure}[tb]
        \begin{subfigure}[b]{0.5\linewidth}
            \begin{minted}[linenos, numbersep=3pt, breaklines, baselinestretch=0.9, escapeinside=||]{Go}
func main() {
    a := make(chan struct{})
    b := make(chan struct{})
    c := make(chan struct{})
    go func() {
        <- a |\label{lst:resolution_example_go_dl1}|
        b <- struct{}{}
        <- c
    }()
    c <- struct{}{} |\label{lst:resolution_example_go_dl2}|
    <- b
    a <- struct{}{}
}
            \end{minted}
            \captionsetup{width=\linewidth}
            \caption{Source Go code}
            \label{lst:resolution_example_go}
        \end{subfigure}
        \begin{subfigure}[b]{0.5\linewidth}
            \[
                \prf{a}{\prf{\bar{b}}{\prf{c}{\inert}}} \paralS \prf{\bar{c}}{\prf{b}{\prf{\bar{a}}{\inert}}}
            \]
            \captionsetup{width=\linewidth}
            \caption{CCS abstraction}
            \label{lst:resolution_example_CCS}
        \end{subfigure}
    \captionsetup{width=\linewidth}
    \caption{Faulty Go code.}
    \label{lst:faulty_examples}
\end{figure}
The first strategy parallelises the blocking actions $a$ and $\bar{c}$, producing the process in \Cref{lst:resolution_example_1_ccs}. The second strategy treats input and output actions asymmetrically: outputs, such as $\bar{c}$, follow the first strategy, while for inputs, such as $a$, a complementary output is placed in parallel at the input level. This preserves the action sequence, leaving the subprocess $\prf{a}{\prf{\bar{b}}{\prf{c}{\inert}}}$ unchanged (\autoref{lst:resolution_example_2_ccs}). Refactoring the Go code breaks the cyclic dependency, resolving the deadlock.

\begin{figure}[tbp]
    \begin{subfigure}{0.5\linewidth}
        \begin{center}
    		\begin{subfigure}[t]{\linewidth}
        		\begin{minted}[linenos, numbersep=3pt, breaklines, baselinestretch=0.9, escapeinside=||]{Go}
func main() {
    a := make(chan struct{})
    b := make(chan struct{})
    c := make(chan struct{})
    go func() {
|\removed{- \qquad\quad <- a}|
|\added{+ \qquad\quad go func()\{<- a\}()}|
        b <- struct{}{}
        <- c
    }()
|\removed{- \quad c <- struct\{\}\{\}}|
|\added{+ \quad go func()\{c <- struct\{\}\{\}\}()}|
    <- b
    a <- struct{}{}
}
                \end{minted}
                \captionsetup{width=\linewidth}
                \caption{Resolved Go code (first strategy)}
                \label{lst:resolution_example_1_go}
            \end{subfigure}
    		\begin{subfigure}[t]{\linewidth}
    			\[
    				( \prf{a}{\inert} \paralS \prf{\bar{b}}{\prf{c}{\inert}} ) \paralS ( \prf{\bar{c}}{\inert} \paralS \prf{b}{\prf{\bar{a}}{\inert}} )
    			\]
    			\captionsetup{width=\linewidth}
    			\caption{Resolved CCS form (first strategy)}
    			\label{lst:resolution_example_1_ccs}
    		\end{subfigure}
        \end{center}
    \end{subfigure}
    \begin{subfigure}{0.5\linewidth}
        \begin{center}
    		\begin{subfigure}[t]{\linewidth}
    			\begin{minted}[linenos, numbersep=3pt, breaklines, baselinestretch=0.9, escapeinside=||]{Go}
func main() {
    a := make(chan struct{})
    b := make(chan struct{})
    c := make(chan struct{})
    go func() {
|\added{+ \qquad\quad go func()\{a <- struct\{\}\{\}\}()}|
        <- a
        b <- struct{}{}()
        <- c
    }()
|\removed{- \quad c <- struct\{\}\{\}}|
|\added{+ \quad go func()\{c <- struct\{\}\{\}\}()}|
    <- b
|\removed{- \quad a <- struct\{\}\{\}}|
}
    			\end{minted}
    			\captionsetup{width=\linewidth}
    			\caption{Resolved Go code (second strategy)}
    			\label{lst:resolution_example_2_go}
    		\end{subfigure}
    		\begin{subfigure}[t]{\linewidth}
    			\[
    		 		( \prf{\bar{a}}{\inert} \paralS \prf{a}{\prf{\bar{b}}{\prf{c}{\inert}}} ) \paralS ( \prf{\bar{c}}{\inert} \paralS \prf{b}{\inert} )
    			\]
    			\captionsetup{width=\linewidth}
    			\caption{Resolved CCS form (second strategy)}
    			\label{lst:resolution_example_2_ccs}
    		\end{subfigure}
        \end{center}
    \end{subfigure}
    \captionsetup{width=\linewidth}
    \caption{Fixed Go code}
    \label{lst:resolution_examples}
\end{figure}

\subsection{Important Go Patterns}\label{sec:important_go_pat}
Some Go patterns require extra attention during the analysis because they can significantly alter the behaviour of a program, including deadlock status.

\subsubsection{Channel closing in Go}
can enable blocked programs to terminate: reading from a closed channel succeeds immediately and returns the zero value of the channel's type. For instance, \Cref{code:block_read} deadlocks on an unfulfilled read, while \Cref{code:nonblock_read} terminates because the closed channel allows the read to return the default \verb|int| value. Conversely, writing to or re-closing a closed channel triggers a runtime panic, as shown in \Cref{code:closed_write,code:double_close}.

\begin{figure}[t]
    \begin{subfigure}{0.5\textwidth}
        \begin{minted}[linenos, numbersep=3pt, breaklines, baselinestretch=0.9, escapeinside=||]{Go}
func main() {
    ch := make(chan int)
    <-ch
}
        \end{minted}
        \caption{Blocking read}
        \label{code:block_read}
    \end{subfigure}
    \begin{subfigure}{0.5\textwidth}
        \begin{minted}[linenos, numbersep=3pt, breaklines, baselinestretch=0.9, escapeinside=||]{Go}
func main() {
    ch := make(chan int)
    close(ch)
    <-ch
}
        \end{minted}
        \caption{Non-blocking read}
        \label{code:nonblock_read}
    \end{subfigure}
    \caption{Channel closing in Go}
    \label{code:channel_closing}
\end{figure}

\begin{figure}[t]
    \begin{subfigure}{0.5\textwidth}
        \begin{minted}[linenos, numbersep=3pt, breaklines, baselinestretch=0.9, escapeinside=||]{Go}
func main() {
    ch := make(chan int)
    close(ch)
    ch <- 42
}
        \end{minted}
        \caption{Closed write}
        \label{code:closed_write}
    \end{subfigure}
    \begin{subfigure}{0.5\textwidth}
        \begin{minted}[linenos, numbersep=3pt, breaklines, baselinestretch=0.9, escapeinside=||]{Go}
func main() {
    ch := make(chan int)
    close(ch)
    close(ch)
}
        \end{minted}
        \caption{Double close}
        \label{code:double_close}
    \end{subfigure}
    \caption{Channel misuse runtime errors}
    \label{code:runtime_errors}
\end{figure}

\NeoGDDR maintains channel attributes via an extensible key-value \textit{struct}, including a boolean indicating if a channel is closed. During analysis, the algorithm references these attributes to determine whether an action should trigger a reduction or report an error. 

While the baseline algorithm considers any unsynchronised action a deadlock, channel state modifies this behaviour. For read operations, a closed channel enables unconditional reduction, whereas an open channel requires synchronous matching. Conversely, write operations on closed channels are logged as errors; operations on open channels remain unchanged.

\subsubsection{Asynchronous Channels}
utilise a queue with a fixed capacity defined at creation. This buffer allows communication to proceed without blocking under specific conditions:
\begin{itemize}
    \item Reads are non-blocking if the buffer contains values.
    \item Writes are non-blocking if the buffer has available capacity.
\end{itemize}
Conversely, a read on an empty buffer or a write to a full buffer causes the goroutine to block until a complementary action occurs, mirroring synchronous behaviour. For example, the unbuffered write in~\Cref{code:block_write} deadlocks, whereas the buffered channel in~\Cref{code:buf_write} terminates normally.

\begin{figure}[t]
    \begin{subfigure}{0.5\textwidth}
        \begin{minted}[linenos, numbersep=3pt, breaklines, baselinestretch=0.9, escapeinside=||]{Go}
func main() {
    ch := make(chan int)
    ch <- 42
}
        \end{minted}
        \caption{Synchronous channel write}
        \label{code:block_write}
    \end{subfigure}
    \begin{subfigure}{0.5\textwidth}
        \begin{minted}[linenos, numbersep=3pt, breaklines, baselinestretch=0.9, escapeinside=||]{Go}
func main() {
    ch := make(chan int, 5)
    ch <- 42
}
        \end{minted}
        \caption{Asynchronous channel write}
        \label{code:buf_write}
    \end{subfigure}
    \caption{Channel writing}
    \label{code:channel_writing}
\end{figure}

By adding capacity to the channel attributes, the algorithm handles communication as follows:
\begin{itemize}
    \item If capacity is 0, the channel is treated as synchronous.
    \item If the buffer is full, reads reduce immediately, while writes are treated as synchronous.
    \item If the buffer is empty, writes are reduced immediately, while reads are treated as synchronous.
    \item Otherwise, any action is unconditionally reduced.
\end{itemize}
By updating the buffer occupancy during these transitions, asynchronous operations either complete immediately or default to existing synchronous logic.

\subsubsection{Recursion.}
To prevent infinite loops during CCS generation, recursive function definitions are stored and expanded dynamically during analysis. However, infinite execution paths in the source program can still cause the symbolic analysis to diverge. To resolve this, \NeoGDDR caches explored process states; if a state recurs, the redundant path is truncated. 

Such loops represent \emph{livelocks}, where threads communicate indefinitely without termination. Since these do not constitute deadlocks, they can be safely omitted. While state caching does not directly identify deadlocks, it enables the analysis of recursive processes, facilitating the detection of deadlocks that would otherwise remain unreachable.



\section{State of the Art}\label{sec:stateoftheart}

\subsection{Deadlock Detection Evaluation}
To evaluate \NeoGDDR's detection capabilities, we benchmarked it against state-of-the-art tools: GCatch~\cite{GCatch}, \goat~\cite{Goat}, Gomela~\cite{Gomela}, and Gong~\cite{Gong}. Due to varying development timelines and methodologies, dependency compatibility across Go versions was inconsistent. Testing across versions from Go 1.16 onward identified Go 1.18~\cite{Go1_18} as the most stable environment, satisfying the majority of requirements and yielding the most reliable results. The only exception is Gomela, which utilises a pre-configured Docker container running Go 1.17.

\todo[inline]{Benchmark extended to 25 cases, with the source of each new one.}
We evaluated the tools using the \verb|go-deadlock-bug-collection| benchmark~\cite{BugCollection,GoDDaR}. This collection was extended to 25 test cases with programs from other benchmarks: \verb|preload|, \verb|gomela-circular-dependency|, \verb|gomela-file-parser|, and \verb|gomela-x-sender-x-receiver-plus-1| from Gomela; \verb|nested|, \verb|select-with-continuation|, and \verb|serving_2137| from \goat; and \verb|cockroachdb_2448|, \verb|grpc-go_660|, and \\\verb|kubernetes_70277| from GoKer, the bug-kernel suite of GoBench~\cite{GoBench}. Each test case consists of a real-world bug reduced to its minimal reproducible core, typically under~50 lines of code (the largest, \verb|serving_2137|, has 143). These bugs include deadlocks, double channel closures, and other channel-related runtime exceptions.

Detailed results are presented in \Cref{tab:neogoddar_table}. \Cref{tab:tool_stats} summarises these findings, showing the total true positives, false negatives, and failed analyses for each tool.

\todo[inline]{Table updated with the new benchmark results and the execution time (s) of each tool.}
\newcommand{\timed}[2]{\shortstack{#1\\[-1pt]{\tiny #2\,s}}}
\begin{table}[htbp]
\setlength{\aboverulesep}{0pt}
\setlength{\belowrulesep}{0pt}
\renewcommand{\arraystretch}{2} 
    \centering\scriptsize
    \caption{Deadlock detection benchmark.}
    \label{tab:neogoddar_table}
    \rowcolors{2}{black!10}{}
    \setlength{\tabcolsep}{6pt}
    \begin{tabular}{l*5{c}}
        \toprule
        \textbf{Bug Id}  & \textbf{GCatch} & \textbf{\goat} & \textbf{Gomela} & \textbf{Gong} & \textbf{\NeoGDDR} \\
        \midrule
        \verb|cockroachdb_13755| & \timed{\xmark}{2.55} & \timed{\xmark}{0.74} & \timed{\xmark}{1.25} & \timed{C}{0.47} & \timed{C}{0.58} \\
        \midrule[0.15pt]
        \verb|cockroachdb_18101| & \timed{\xmark}{2.00} & \timed{\xmark}{1.01} & \timed{\cmark}{6.36} & \timed{C}{0.47} & \timed{\cmark}{0.66} \\
        \midrule[0.15pt]
        \verb|cockroachdb_2448| & \timed{\xmark}{2.16} & \timed{\cmark}{1.07} & \timed{\cmark}{2.80} & \timed{C}{0.49} & \timed{\cmark}{0.65} \\
        \midrule[0.15pt]
        \verb|cockroachdb_24808| & \timed{\xmark}{0.52} & \timed{\xmark}{0.13} & \timed{\xmark}{0.58} & \timed{C}{0.45} & \timed{\cmark}{0.13} \\
        \midrule[0.15pt]
        \verb|cockroachdb_25456| & \timed{\cmark}{0.58} & \timed{\cmark}{0.14} & \timed{C}{0.60} & \timed{\cmark}{0.47} & \timed{\cmark}{0.13} \\
        \midrule[0.15pt]
        \verb|etcd_6857| & \timed{\cmark}{0.61} & \timed{\cmark}{0.13} & \timed{C}{0.60} & \timed{C}{0.43} & \timed{\cmark}{0.12} \\
        \midrule[0.15pt]
        \verb|gomela-circular-dependency| & \timed{\cmark}{0.57} & \timed{\cmark}{0.13} & \timed{\cmark}{1.82} & \timed{C}{0.46} & \timed{\cmark}{0.12} \\
        \midrule[0.15pt]
        \verb|gomela-file-parser| & \timed{\xmark}{0.50} & \timed{\cmark}{0.13} & \timed{\cmark}{4.20} & \timed{C}{0.43} & \timed{\cmark}{0.13} \\
        \midrule[0.15pt]
        \verb|gomela-x-sender-x-receiver-plus-1| & \timed{\xmark}{0.56} & \timed{\xmark}{0.13} & \timed{\cmark}{1.82} & \timed{C}{0.45} & \timed{\cmark}{0.13} \\
        \midrule[0.15pt]
        \verb|grpc-go_1275| & \timed{\cmark}{0.55} & \timed{\cmark}{0.13} & \timed{\cmark}{1.78} & \timed{\cmark}{0.41} & \timed{\cmark}{0.12} \\
        \midrule[0.15pt]
        \verb|grpc-go_1424| & \timed{\cmark}{0.56} & \timed{\cmark}{0.13} & \timed{\cmark}{1.74} & \timed{\cmark}{0.43} & \timed{\cmark}{0.12} \\
        \midrule[0.15pt]
        \verb|grpc-go_490| & \timed{\xmark}{2.21} & \timed{\xmark}{0.77} & \timed{\xmark}{1.21} & \timed{C}{0.43} & \timed{\cmark}{0.60} \\
        \midrule[0.15pt]
        \verb|grpc-go_660| & \timed{\cmark}{0.60} & \timed{C}{0.13} & \timed{C}{0.58} & \timed{C}{0.43} & \timed{C}{0.13} \\
        \midrule[0.15pt]
        \verb|kubernetes_25331| & \timed{\xmark}{2.16} & \timed{\xmark}{1.05} & \timed{\xmark}{2.73} & \timed{C}{0.45} & \timed{C}{0.65} \\
        \midrule[0.15pt]
        \verb|kubernetes_35672| & \timed{\cmark}{0.57} & \timed{\cmark}{0.13} & \timed{\cmark}{1.88} & \timed{\cmark}{0.43} & \timed{\cmark}{0.12} \\
        \midrule[0.15pt]
        \verb|kubernetes_5316| & \timed{\cmark}{2.30} & \timed{\cmark}{1.14} & \timed{\cmark}{2.52} & \timed{\xmark}{0.43} & \timed{\cmark}{0.52} \\
        \midrule[0.15pt]
        \verb|kubernetes_70277| & \timed{\xmark}{1.95} & \timed{\cmark}{1.09} & \timed{\cmark}{3.28} & \timed{C}{0.46} & \timed{C}{0.51} \\
        \midrule[0.15pt]
        \verb|moby_21233| & \timed{\xmark}{1.65} & \timed{\cmark}{0.92} & \timed{C}{1.16} & \timed{C}{0.43} & \timed{C}{0.58} \\
        \midrule[0.15pt]
        \verb|moby_33293| & \timed{\cmark}{0.56} & \timed{\cmark}{0.13} & \timed{C}{0.60} & \timed{C}{0.43} & \timed{\cmark}{0.11} \\
        \midrule[0.15pt]
        \verb|moby_33781| & \timed{\cmark}{1.01} & \timed{\cmark}{0.16} & \timed{\cmark}{2.67} & \timed{\xmark}{0.59} & \timed{\cmark}{0.15} \\
        \midrule[0.15pt]
        \verb|moby_4395| & \timed{\cmark}{0.54} & \timed{\cmark}{0.13} & \timed{C}{0.61} & \timed{\cmark}{0.42} & \timed{\cmark}{0.12} \\
        \midrule[0.15pt]
        \verb|nested| & \timed{\cmark}{0.58} & \timed{\cmark}{0.13} & \timed{\cmark}{1.75} & \timed{\cmark}{0.47} & \timed{\cmark}{0.12} \\
        \midrule[0.15pt]
        \verb|preload| & \timed{\xmark}{1.54} & \timed{\xmark}{0.92} & \timed{\cmark}{40.55} & \timed{C}{0.47} & \timed{\cmark}{1.02} \\
        \midrule[0.15pt]
        \verb|select-with-continuation| & \timed{\cmark}{0.53} & \timed{\cmark}{0.14} & \timed{\cmark}{1.86} & \timed{\cmark}{0.44} & \timed{\cmark}{0.12} \\
        \midrule[0.15pt]
        \verb|serving_2137| & \timed{\xmark}{1.46} & \timed{\cmark}{1.02} & \timed{C}{1.11} & \timed{C}{0.48} & \timed{C}{0.40} \\
        \bottomrule
    \end{tabular}
\end{table}

\todo[inline]{Table updated with the new results; the Invalid row is now Crashes.}
\begin{table}[tb]
\setlength{\aboverulesep}{0pt}
\setlength{\belowrulesep}{0pt}
\renewcommand{\arraystretch}{1.45} 
\centering\scriptsize
    \caption{Tool statistics summary.}
    \label{tab:tool_stats}
    \rowcolors{2}{black!10}{}
    \begin{tabular}{l*5{>{~~~}c}}
        \toprule
        \textbf{Mark}                & \textbf{GCatch} & \textbf{\goat} & \textbf{Gomela} & \textbf{Gong} & \textbf{\NeoGDDR} \\
        \midrule
        True positives (\cmark)      & 13              & 17             & 14              & 7             & 19 \\
        \midrule[0.15pt]
        False negatives (\xmark)     & 12              & 7              & 4               & 2             & 0 \\
        \midrule[0.15pt]
        Crashes (C)                  & 0               & 1              & 7               & 16            & 6 \\
        \bottomrule
    \end{tabular}
\end{table}


\todo[inline]{Legend reduced to \cmark, \xmark and C; the test-case property marks were removed.}
Tool performance is categorised by the following indicators:
\begin{itemize}
    \item \cmark: Tool gives the expected result (correctly identified the bug)
    \item \xmark: Tool gives wrong result (false negative)
    \item C: Tool exception/crash.
\end{itemize}

\todo[inline]{Removed the moby\_33293 example, which no longer fails.}
Gong and \NeoGDDR are the only evaluated tools that use \migo as an intermediate representation (IR); consequently, both are limited by \migoinf's ability to generate valid IR, like the cases it crashes. Nevertheless, \NeoGDDR outperforms Gong by addressing extra communication patterns like those discussed in \Cref{sec:important_go_pat}.

\todo[inline]{Comparison rewritten with the new results (\NeoGDDR finds the most bugs; \goat has better coverage).}
Compared to the other tools (see~\Cref{tab:tool_stats}), \NeoGDDR reports the most true positives (19 of 25) and is the only tool with no false negatives. Despite the limitations of \migoinf during IR generation, \NeoGDDR failed to analyse fewer programs (6) than Gong (16), the other \migo-based tool. GCatch completed all analyses without crashing, and \goat crashed on a single case, but both missed many of the bugs they analysed: GCatch has 12 false negatives and \goat has 7.
%
%
\NeoGDDR and \goat are the strongest tools in the benchmark. \NeoGDDR identifies two more bugs (19 against 17) and has no false negatives (against 7 for \goat), while \goat has far better coverage, analysing 24 of 25 programs against 19 for \NeoGDDR. This is expected: \goat's near-exhaustive control-flow graph analysis does not depend on \migoinf, whereas \NeoGDDR's crashes are largely bounded by that intermediate representation.

\todo[inline]{Size discussion rewritten for the new sizes and times.}
\Cref{tab:ccs_program_sizes} provides a quantitative measure of the complexity of the benchmark test cases by detailing the lines of code (LoC) of the source files and the “size” of the generated CCS processes for each program.%
\footnote{``Size'' is the number of nodes of the AST representing the CCS term.}
Notice that the Go code considered corresponds to minimal examples showcasing the deadlock when an issue was opened (so, naturally, they are small programs).
This size metric reflects the scale of 
CCS term that \NeoGDDR must explore during symbolic analysis. The data reveals that the programs that could be translated result in small CCS programs: sizes range from~3 to~67, with a median of~25, and 16 of the 19 are below~40. The largest program analysed is \verb|cockroachdb_2448|, with a size of~67, which \NeoGDDR analysed in~0.65 seconds (see \Cref{tab:neogoddar_table}). Additionally, the table explicitly identifies instances of “parse failure,” which occur when the external migoinfer tool is unable to generate a valid MiGo intermediate representation, a known limitation that accounts for the “invalid” results noted in the detection evaluation.

\todo[inline]{Table recomputed for all 25 cases: non-blank LoC, CCS nodes (leaves included), parse failure where the tool crashed.}
\begin{table}[tb]
\setlength{\aboverulesep}{0pt}
\setlength{\belowrulesep}{0pt}
\renewcommand{\arraystretch}{1.45} 
\centering\scriptsize
    \caption{CCS program sizes}
    \label{tab:ccs_program_sizes}
    \rowcolors{2}{black!10}{}
    \begin{tabular}{l*1{>{~~~}c}*1{>{~~~}c}}
        \toprule
        \textbf{File} & \textbf{Go LoC} & \textbf{CCS Program Size} \\
        \midrule
        \verb|cockroachdb_13755/main.go|          & 42 & parse failure \\
        \midrule[0.15pt]
        \verb|cockroachdb_18101/main.go|          & 22 & 34 \\
        \midrule[0.15pt]
        \verb|cockroachdb_2448/main.go|           & 62 & 67 \\
        \midrule[0.15pt]
        \verb|cockroachdb_24808/main.go|          & 40 & 22 \\
        \midrule[0.15pt]
        \verb|cockroachdb_25456/main.go|          & 14 & 3 \\
        \midrule[0.15pt]
        \verb|etcd_6857/main.go|                  & 58 & 48 \\
        \midrule[0.15pt]
        \verb|gomela-circular-dependency/main.go| & 13 & 27 \\
        \midrule[0.15pt]
        \verb|gomela-file-parser/main.go|         & 16 & 33 \\
        \midrule[0.15pt]
        \verb|gomela-x-sender-x-receiver-plus-1/main.go| & 17 & 33 \\
        \midrule[0.15pt]
        \verb|grpc-go_1275/main.go|               & 24 & 6 \\
        \midrule[0.15pt]
        \verb|grpc-go_1424/main.go|               & 12 & 3 \\
        \midrule[0.15pt]
        \verb|grpc-go_490/main.go|                & 24 & 9 \\
        \midrule[0.15pt]
        \verb|grpc-go_660/main.go|                & 26 & parse failure \\
        \midrule[0.15pt]
        \verb|kubernetes_25331/main.go|           & 53 & parse failure \\
        \midrule[0.15pt]
        \verb|kubernetes_35672/main.go|           & 26 & 25 \\
        \midrule[0.15pt]
        \verb|kubernetes_5316/main.go|            & 30 & 35 \\
        \midrule[0.15pt]
        \verb|kubernetes_70277/main.go|           & 31 & parse failure \\
        \midrule[0.15pt]
        \verb|moby_21233/main.go|                 & 31 & parse failure \\
        \midrule[0.15pt]
        \verb|moby_33293/main.go|                 & 14 & 11 \\
        \midrule[0.15pt]
        \verb|moby_33781/main.go|                 & 21 & 23 \\
        \midrule[0.15pt]
        \verb|moby_4395/main.go|                  & 24 & 9 \\
        \midrule[0.15pt]
        \verb|nested/main.go|                     & 16 & 18 \\
        \midrule[0.15pt]
        \verb|preload/main.go|                    & 41 & 64 \\
        \midrule[0.15pt]
        \verb|select-with-continuation/main.go|   & 17 & 36 \\
        \midrule[0.15pt]
        \verb|serving_2137/main.go|               & 143 & parse failure \\
        \bottomrule
        \rowcolor{white}\multicolumn{3}{l}{The parse failures are reported as tool crash in \Cref{tab:neogoddar_table}. }
    \end{tabular}
\end{table}


\subsection{Deadlock Resolution Evaluation}
To evaluate our deadlock resolution approach, we present five representative deadlocks in~\Cref{lst:resolution_eval_examples}. We compare the fixes provided by \NeoGDDR's two strategies against GFix~\cite{GCatch}, the resolution component of GCatch. The evaluation results are summarised in~\Cref{tab:resolution_evaluation}.

Sequential~(\autoref{lst:resolution_sequential}) represents a case where two dual communication operations are placed sequentially in the same goroutine. As shown in~\autoref{lst:resolution_sequential_fixed}, \NeoGDDR parallelises the first communication to enable synchronisation.
Circular~(\autoref{lst:resolution_circular}) denotes a deadlock caused by a circular dependency between two channels; our tool resolves this by parallelising the send operation on channel \texttt{b}~(\autoref{lst:resolution_circular_fixed}).
Triple refers to the example in~\autoref{lst:resolution_example_go}, where communication operations across three channels are incorrectly ordered.
Select~(\autoref{lst:resolution_select}) involves a cyclic dependency within a \texttt{select} statement branch, which \NeoGDDR fixes by parallelising the receive on channel~\texttt{c}~(\autoref{lst:resolution_select_fixed}).


For these four cases, GFix fails to provide a resolution because they do not align with its targeted deadlock patterns. Conversely, \NeoGDDR resolves them by parallelising blocked operations to unblock the synchronisation chain.
MissComm~(\autoref{lst:resolution_missComm}) represents a missing-communication deadlock where an \texttt{if} statement branch fails to receive a value sent by a second goroutine. By design, \NeoGDDR does not resolve this type of deadlock because such cases often cannot be fixed without altering the program's intended semantics. For instance, while adding a dummy write could unblock a lone reader, the synthesised data might be invalid for the rest of the program's logic. Consequently, we chose not to handle these cases as they typically require manual, context-specific fixes. GFix, however, resolves this bug using its internal strategies, with the first strategy's result shown in~\autoref{lst:resolution_missComm_fixed}.

\begin{table}[tbp]
\centering
\caption{Comparison of deadlock resolution.
 	In column Type, MC denotes a missing communication, and IO denotes incorrect communication ordering.
    S-I, S-II, and S-III, are the tool's deadlock resolution strategies.
}
\label{tab:resolution_evaluation}
\renewcommand\theadfont{\bfseries}
\setlength{\tabcolsep}{1ex} 
\rowcolors{2}{black!10}{}
\begin{tabular}{l*{10}{c}}
\midrule
%
\thead[l]{Bug} & & & & \multicolumn{2}{l}{\thead{\NeoGDDR}} && \multicolumn{3}{l}{\thead{GFix~\cite{GCatch}}}\\[-1.5ex]
\thead[l]{Category} & \thead{List.} & \thead{Type} & & \thead{S-I}      & \thead{S-II}     &&\thead{S-I}    & \thead{S-II}    & \thead{S-III} \\

\midrule
Sequential      &  \ref{lst:resolution_sequential}  & ICO        &   & \cmark          & \cmark          & &\xmark        & \xmark         & \xmark       \\
Circular        &  \ref{lst:resolution_circular}    & ICO        &   & \cmark          & \cmark          & &\xmark        & \xmark         & \xmark       \\
Triple          &  \ref{lst:resolution_example_go}  & ICO        &   & \cmark          & \cmark          & &\xmark        & \xmark         & \xmark       \\
Select          &  \ref{lst:resolution_select}      & ICO        &   & \cmark          & \cmark          && \xmark        & \xmark         & \xmark       \\
MissComm        &  \ref{lst:resolution_missComm}    & MC         &  & \xmark          & \xmark          && \cmark        & \cmark         & \cmark       \\
\bottomrule
\end{tabular}
\end{table}

\newcommand{\figidth}{0.5\linewidth}
\begin{listing}[tbp]
		\begin{subfigure}[b]{\figidth}
			\begin{minted}[linenos, numbersep=3pt, breaklines, baselinestretch=0.9, escapeinside=||]{Go}
func main() {
	a := make(chan struct{})
	<- a
	a <- struct{}{}
}
			\end{minted}
		\captionsetup{justification=centering}
			\caption{Sequential}
			\label{lst:resolution_sequential}
		\end{subfigure}
		\begin{subfigure}[b]{\figidth}
			\begin{minted}[linenos, numbersep=3pt, breaklines, baselinestretch=0.9, escapeinside=||]{Go}
func main() {
    a := make(chan struct{})
|\removed{- \quad <-a}| 	
|\added{+ \quad go func()\{<-a\}()}| 	
    a <- struct{}{}
}
			\end{minted}
			\captionsetup{justification=centering}
			\caption{Sequential Fixed (\NeoGDDR, S-I)}
			\label{lst:resolution_sequential_fixed}
		\end{subfigure}
    \par\bigskip
		\begin{subfigure}[b]{\figidth}
			\begin{minted}[linenos, numbersep=3pt, breaklines, baselinestretch=0.9, escapeinside=||]{Go}
func main() {
	a := make(chan struct{})
	b := make(chan struct{})
	go func() {
		a <- struct{}{}
		<- b
	}()
	b <- struct{}{}
	<- a
}
			\end{minted}
			\captionsetup{justification=centering}
			\caption{Circular}
			\label{lst:resolution_circular}
		\end{subfigure}
		\begin{subfigure}[b]{\figidth}
			\begin{minted}[linenos, numbersep=3pt, breaklines, baselinestretch=0.9, escapeinside=||]{Go}
func main() {
    a := make(chan struct{})
    b := make(chan struct{})
    go func() {
        a <- struct{}{}
        <- b
    }()
|\removed{- \quad b <- struct\{\}\{\}}|
|\added{+ \quad go func()\{b <- struct\{\}\{\}\}()}|
    <- a
}
			\end{minted}
			\captionsetup{justification=centering}
			\caption{Circular Fixed (\NeoGDDR, S-I)}
			\label{lst:resolution_circular_fixed}
		\end{subfigure}
    \par\bigskip
		\begin{subfigure}[b]{\figidth}
			\begin{minted}[linenos, numbersep=3pt, breaklines, baselinestretch=0.9, escapeinside=||]{Go}
func main() {
	a := make(chan struct{})
	b := make(chan struct{})
	c := make(chan struct{})
	go func() {
		a <- struct{}{}
		b <- struct{}{}
		c <- struct{}{}
	}()
	select {
	case <- a:
		<- c
		<- b
	case <- b:
		return
	}
}
            	\end{minted}
			\captionsetup{justification=centering}
			\caption{Select}
			\label{lst:resolution_select}
		\end{subfigure}
		\begin{subfigure}[b]{\figidth}
			\begin{minted}[linenos, numbersep=3pt, breaklines, baselinestretch=0.9, escapeinside=||]{Go}
func main() {
	a := make(chan struct{})
	b := make(chan struct{})
	c := make(chan struct{})
	go func() {
		a <- struct{}{}
		b <- struct{}{}
		c <- struct{}{}
	}()
	select {
	case <- a:
oved{-\qquad\qquad <-c}|
ed{+\qquad\qquad go func()\{<-c\}()}|
		<- b
	case <- b:
		return
	}
}
            	\end{minted}
			\captionsetup{justification=centering}
			\caption{Select Fixed (\NeoGDDR, S-I)}
			\label{lst:resolution_select_fixed}
		\end{subfigure}
    \par\bigskip
		\begin{subfigure}[b]{\figidth}
			\begin{minted}[linenos, numbersep=3pt, breaklines, baselinestretch=0.9, escapeinside=||]{Go}
func main() {
	c := make(chan struct{})
	go func() { // Do work
		c <- struct{}{}
	}()
	if true {
		<- c
	} else {
		return
	}
}
			\end{minted}
			\captionsetup{justification=centering}
			\caption{Missing communication}
			\label{lst:resolution_missComm}
		\end{subfigure}
		\begin{subfigure}[b]{\figidth}
			\begin{minted}[linenos, numbersep=3pt, breaklines, baselinestretch=0.9, escapeinside=||]{Go}
func main() {
|\removed{- c := make(chan struct\{\})}|
|\added{+ c := make(chan struct\{\}, 1)}|
	go func() { // Do work
		c <- struct{}{}
	}()
	if true {
		<- c
	} else {
		return
	}
}
			\end{minted}
			\captionsetup{justification=centering}
			\caption{Missing communication (GFix, S-I)}
			\label{lst:resolution_missComm_fixed}
		\end{subfigure}
	\caption{Deadlock resolution examples.}
	\label{lst:resolution_eval_examples}
\end{listing}

In this paper, we introduce \NeoGDDR, a tool for the detection and resolution of deadlocks in Go. Our approach extracts a communication model from a Go program as a CCS process and uses symbolic execution to identify deadlocks. When a deadlock is detected, the tool provides a suggested fix for the original Go source code. 

\NeoGDDR shows substantial potential, correctly identifying deadlocks in cases where existing tools fail. Our extended CCS, which distinguishes between internal and external choices, effectively models Go's communication semantics. Furthermore, the inclusion of channel closure, asynchronous buffers, and recursion proved critical for accuracy, as \NeoGDDR outperformed other tools utilising the same intermediate representation (\migo). Overall, benchmarks indicate that \NeoGDDR exceeds most state-of-the-art alternatives in both detection coverage (\Cref{tab:tool_stats}) and resolution capability (\Cref{tab:resolution_evaluation}).

This pragmatic approach is validated by our proof-of-concept implementation. However, formal soundness results for the detection and resolution algorithms remain subjects of future work. A key challenge lies in defining and evaluating the “minimal” program modification required to resolve a deadlock while preserving original semantics.


\todo[inline]{Future work now proposes transpiling directly to CCS.}
A significant current limitation is the dependence on \migo, which is no longer actively maintained.
Most failed analyses were caused by the inability of \migoinf to generate valid IR.
Future work will explore alternative representations, such as those used by Gomela, where the IR is transpiled by the authors of the tool.
In Gomela's case, it uses Promela, we would transpile directly to CCS to improve precision, coverage, and independence from external tools.
In the near future we plan to assess the deadlock correction algorithms through a systematic and exhaustive application to all the programs in the detection benchmark.


We also identified an edge case in our recursion handling. In scenarios such as \verb|moby#21233|, the algorithm may enter an infinite loop because duplicate state checks occur only between reductions. Non-reduction steps, such as process calls, can currently repeat indefinitely. Refining this check to include all state transitions is a priority.

Finally, while \NeoGDDR currently focuses on channel-based communication, we intend to extend the analysis to other Go primitives, such as \texttt{sync.Mutex} and \texttt{sync.WaitGroup}. Expanding the range of supported concurrency primitives will enhance the tool's utility for complex, real-world applications.

%

%
%
\bibliographystyle{splncs04}
\bibliography{mybibliography}

\end{document}